\documentstyle[aps,preprint]{revtex}
\tightenlines
\begin{document}
\preprint{DFTUZ/97-26, USACH 97/11}

\title{ A Nambu-Jona-Lasinio like model 
from QCD at low energies}
\author{Jos\'e~Luis~Cort\'es$^\dagger$\thanks{E-mail:
cortes@posta.unizar.es},
Jorge~Gamboa$^\ddagger$\thanks{E-mail: jgamboa@lauca.usach.cl}, 
and Luis Vel\'azquez$^{\ast}$\thanks{E-mail: velazquez@posta.unizar.es}}  
\address{$^\dagger$ Departamento de F\'{\i}sica Te\'orica, 
Universidad de Zaragoza, 50009 Zaragoza, Spain.\\ 
$^\ddagger$Departamento de F\'{\i}sica,Universidad de Santiago de
Chile, Casilla 307, Santiago, Chile. \\
$^{\ast}$ Departamento de Matem\'atica Aplicada, Universidad de Zaragoza,
50015 Zaragoza, Spain.}

\maketitle \date{\today}
\maketitle 
\begin{abstract}
A generalization to any dimension of the fermion field transformation
which allows to derive the solution of the massless Schwinger model
in the path integral framework is identified. New arguments based on 
this transformation for a Nambu-Jona-Lasinio (NJL) like model as the 
low energy limit of a gauge theory in dimension greater than two are 
presented. Our result supports the spontaneous chiral symmetry breaking
picture conjectured by Nambu many years ago and the link between QCD, 
NJL and chiral models.

\bigskip\bigskip
\noindent PACS number(s): 12.38-t , 12.38.Aw 
\bigskip
\end{abstract}

\pacs{}
\narrowtext

Although the theory of strong interactions is known already for
a long time our understanding of hadronic physics
is still very limited.  While at high energies a simple description
in terms of quark and gluon degrees of freedom is possible no
similar simple description of the low energy nonperturbative 
domain is still available and the basic mechanisms behind the
low energy properties (confinement and chiral symmetry
breaking) are not well understood. A candidate for a model
incorporating dynamical symmetry breaking and goldstone bosons is
the Nambu-Jona-Lasinio
model~\cite{NJL} as a bridge between the high energy description in
terms of quark and gluons and the chiral Lagrangian description 
at low energies. The absence of low mass glueballs can also be
used as an argument in favour of an effective fermionic theory at 
intermediate energies. A Nambu-Jona-Lasinio like effective model
of QCD at intermediate energies~\cite{RMP,Kle} provides a new 
framework for studying its non-perturbative behavior. A comparison
of the Nambu-Jona-Lasinio with esperimental data and extensions
are considered in~\cite{BBR} .
 The possibility to use the relation 
of the Nambu-Jona-Lasinio model with QCD in order to make
some progress in the understanding of the low energy hadronic
properties depends on the knowledge of the details of
this (possible) relation. The aim of this work is to present new 
arguments, in the path integral formulation, in favor of a fermionic 
model as an approximation to a gauge theory.

The main idea is to introduce an effective fermionic field such that,
as a consequence of the change of variables, an effective mass for the
gauge field is generated. If one considers the low energy expansion 
of the integration over the gauge field, a Lagrangian with contact 
interactions for the effective field is obtained.

The Schwinger model (QED in $D=2$) is an example where a
transformation of the fermionic field~\cite{Fuj}, generating a 
mass term for the gauge field, is used. The transformation is 
in this case an axial $U(1)$ transformation with a parameter 
identified as the scalar field whose derivatives give the 
transverse part of the gauge field.
A peculiarity of the two dimensional case is that, 
simultaneously to the generation of an effective mass for the
gauge field, the effective fermionic field decouples and the 
model reduces to a free theory~\cite{SM}.
   
An axial $U(1)$ transformation in $D > 2$ is not the required
change of variables since the Jacobian of the 
transformation~\cite{Fuj} will not be linear in the gauge field. 
An alternative is to consider a transformation 
$\psi \to \chi$ where

\begin{eqnarray}
    \psi (x) = \Omega (x) \; \chi (x) & \; \; ,  \; \; &
    \Omega (x) =  \exp [{i \over 2} \omega^{\mu\nu} (x) 
    \gamma_\mu \gamma_\nu]
\label{eff}
\end{eqnarray}

\noindent with $\omega^{\mu\nu} = - \omega^{\nu\mu}$. In two
dimensions $i \gamma^0 \gamma^1 = \gamma^5$ and the change of
variables in (\ref{eff}) reduces to an axial $U(1)$ transformation.

The Jacobian ${\cal J} [\omega, A]$ of the transformation (\ref{eff})  
is a ratio of fermionic determinants,

\begin{equation}
  {\cal J} = {{\det \, [ \, i {\partial \hspace{-0.6em} \slash
  \hspace{0.15em}} + {A \hspace{-.6em} \slash \hspace{.15em}} \, ]} \over 
  {\det \, [ \, \Omega \, ( i {\partial \hspace{-0.6em} \slash
  \hspace{0.15em}} + {A \hspace{-.6em} \slash \hspace{.15em}} \, ) \,
  \Omega \, ]}}
  \;.
\label{Jac}
\end{equation} 

\noindent This result is obtained by comparing the result of the
fermionic integration using the original field $\psi$ as integration
variable with the same integral evaluated in terms of the 
effective fermionic field $\chi$ .

In the evaluation of the fermionic determinants, the gauge field
$A^\mu(x)$ and the parameters $\omega^{\mu\nu}(x)$ are external
fields coupled to the fermionic field through the fermionic part
of the gauge theory Lagrangian which,
written in terms of the field $\chi$, takes the following form:

\begin{equation}
{\cal L}_g \, = \, {\bar \chi} \; \Omega(\omega) 
\, [ \, i {\partial \hspace{-0.6em} \slash \hspace{0.15em}} + {A
\hspace{-.6em} \slash \hspace{.15em}} \, ] \, \Omega(\omega) \; \chi
\;.
\end{equation}

\noindent In the case of an infinitesimal transformation, the 
dominant contribution to the Jacobian can be obtained from the 
one loop diagram with two fermionic propagators connecting
a vertex with a gauge field as external line with another
vertex where the gauge field is replaced by  $\omega^{\mu\nu}$ . 

An ultraviolet regularization is required in order to
have a well defined Jacobian (and field transformation) for
$D > 2$ . A trivial calculation gives

\begin{equation}
{\cal J}[\omega ,A] = 1 + \Lambda^{D-2} \int d^Dx \;
\partial_\nu \omega^{\mu\nu} \; A_\mu + \ldots
\end{equation}

\noindent where $\Lambda$ is an energy scale proportional to
the ultraviolet cut-off with a proportionality factor
depending on the details of the regularization (covariant
momentum cut-off, heat kernel method~\cite{hkm},$\ldots$).
Higher order terms in the Jacobian include corrections 
supressed by factors $\omega$ and/or $1/\Lambda$ .

The next step is to make an appropriate choice of the
effective fermionic field, i.e. a choice for $\omega^{\mu\nu}$ ,
allowing to integrate over the gauge field and to obtain an
effective fermionic model. Given the general decomposition 
of a vector field

\begin{equation}
    A_\mu \; = \; \partial_\mu \; \alpha \; + \;
    \epsilon_{\mu\nu\mu_1 \mu_2 ... \mu_{D-2}} \; \partial^\nu
    \; \beta^{\mu_1 \mu_2 ... \mu_{D-2}} \;,
\label{A}
\end{equation}
 
\noindent it is possible to consider a change of fermionic variables
(\ref{eff}) depending on the gauge field where

\begin{equation}
  \omega_{\mu\nu} \; = \; {1\over 4} \; \epsilon \; 
  \epsilon_{\mu\nu\mu_1 \mu_2 ... \mu_{D-2}} \;
  \beta^{\mu_1 \mu_2 ... \mu_{D-2}} \;.
\label{omega}
\end{equation}

\noindent As a consequence of the Jacobian of the change of 
variables one has, at leading order in $\epsilon$ , an 
additional term in the Lagrangian

\begin{equation}
   {\cal L}_J \; = \; {1\over 4} \; \epsilon \; \Lambda^{D-2} \; A_\mu \;
   ( g^{\mu\nu} - {{\partial^\mu \partial^\nu} \over \Box}) \; A_\nu \;.
\label{LJ}
\end{equation}

\noindent which can be seen as a mass term for the vector field
$A_\mu$ in the Lorentz gauge, $\partial^\mu A_\mu =0$. 

It is a difficult dynamical issue to see whether there is a physical 
mass scale corresponding to a limit $\epsilon \to 0$, 
$\Lambda \to \infty$ with a fixed value for the product
$\epsilon \Lambda^2$ in $D=4$. If this is the case then, in the range 
of energies $E$ such that $E^{2} \ll \epsilon \Lambda^{2}$,
the effect of the gauge interaction reduces to a point-like four-fermion 
coupling and after integration over the gauge field one has a Lagrangian
for the effective fermionic field

\begin{eqnarray}
   {\cal L}_{fer}^{(1)} \; & = & \; 
{\bar \chi} i {\partial \hspace{-0.6em} \slash \hspace{0.15em}} \chi \;
    + 1/(\epsilon \Lambda^{2}) \; \;
    {\bar \chi} \gamma^\mu \chi \; {\bar \chi} \gamma_\mu \chi \nonumber
    \\
    & & + \;  a \; {\bar \chi} \gamma^\mu \chi \; 
    {{\partial_\mu \partial_\nu} 
    \over \Box} \;  {\bar \chi} \gamma^\nu \chi \;.
\label{efL}
\end{eqnarray}
   
\noindent The last term has a coefficient $a$ depending on the 
gauge fixing used in the integration over $A_\mu$ but the gauge 
invariance of the theory guarantees that the $a$-dependence
will cancel at the level of observables. One can take in particular
a gauge fixing such that $a=0$ and the fermionic effective Lagrangian
is just a Thirring model in four dimensions.  

The fermionic system is a non-renormalizable theory that has to be 
understood as an effective theory with a limited energy range of
validity. As any effective theory~\cite{Wein} it will have
a natural scale $M$ (scale of the energy expansion) which allows to 
introduce a dimensionless parameter for each term in the effective
Lagrangian. One has for example a dimensionless effective self-coupling
$g$ associated to the current-current interaction 

\begin{equation}
  g \; = \; {M^2 \over {\epsilon \Lambda^2}} \;.
\end{equation}

\noindent In the derivation of the fermionic Lagrangian (\ref{efL})
the effect of the change of variables at the level of the
Lagrangian has not been taken into account. When the gauge theory
action is written in terms of the effective fermionic field $\chi$
one has an expansion in powers of $\epsilon$ or equivalently
an expansion in $M^2 /\Lambda^2$ . If the scale of the effective 
fermionic theory can be made arbitrarily small in units of the
ultraviolet cut-off of the original gauge theory then the corrections
due to the expansion in $\epsilon$ will be arbitrarily small 
justifying the approximation used to get (\ref{efL}). Note that
as a consequence of the quadratic ultraviolet divergence, the Jacobian 
generates a term in the fermionic Lagrangian that, although proportional
to $\epsilon$, is not suppressed by powers of $M^2/\Lambda^2$ .

The fluctuations for the fermionic field $\chi$ in the 
low energy effective theory are restricted to the energy range
$E < M$. This restriction is crucial for the consistency of the
approximation where the mass term from the Jacobian is retained
while the effect of the change of variables in the Lagrangian is
neglected. In fact if the fermionic field is integrated out directly 
before eliminating the gauge field then the standard fermionic 
determinant of a gauge invariant term has to be reobtained so the 
quadratic term in (\ref{LJ}) has to be canceled with a similar term
from the $\chi$ field integration over fluctuations with $E < M$.
 
\noindent The previous arguments suggest a possible ``derivation'' of 
a fermionic effective theory in a Wilsonian approach as a consequence 
of the integration over energy bands. The result in (\ref{efL})
differs from the generic renormalization group result which would
contain a linear combination of all possible four-fermion terms. This
difference is more striking if one considers several fermionic fields
and/or several gauge fields.
   
In fact all the arguments leading to the fermionic representation of the 
gauge theory can be trivially generalized to the non-abelian $SU(N_c)$ 
case with one or several multiplets of fermionic fields in the 
fundamental representation. One has to include a flavor (I) and a color 
(i) index in the fermionic field. The vector field $A_\mu$ becomes a 
Lie-algebra valued gauge field \mbox{$A_\mu = A_{\mu}^{a} T^a$} and, at 
the same time, one has to consider a fermionic field transformation 
(\ref{eff}) with \mbox{$\omega_{\mu\nu} = \omega_{\mu\nu}^{a} T^{a}$}. 
If the parameter $\omega_{\mu\nu}^{a}$ is obtained from the component 
$A_{\mu}^{a}$ of the gauge field by (\ref{A}) and (\ref{omega}), the 
result for the Jacobian of the transformation will be still given by 
(\ref{LJ}) with a trace over color indices and no derivative terms
if the Lorentz gauge condition $\partial^\mu A_{\mu}^{a} = 0$ is used. 
The four-fermion coupling generated in this case after integration 
over the gauge field is given by 

\begin{equation}
   {\cal L}_{4f}^{(n.a.)} \; = \; 1 /(\epsilon \Lambda^{2})
   \;  {\bar \chi}_I \gamma^\mu T^a \chi_I \;
    {\bar \chi}_J \gamma_\mu T^a \chi_J \;.
\end{equation}

\noindent It can be surprising at first look that the strong gluonic
self-interactions at low energies do not play any role in the 
approximation used for the gauge field integration. Actually this
is not the case because such strong non-perturbative effects are
crucial in the dynamics responsible of the implicitly assumed
dynamical mass $M$. The effects of the gluonic self-interactions
of the non-abelian gauge theory will also be incorporated in the 
effective fermionic theory through the determination of the values of 
the parameters of the effective theory, like the dimensionless 
four-fermion coupling $g$, by the matching of both theories. 

Using the identity 

\begin{equation}
    T_{ij}^a T_{kl}^a \; = \; {1\over 2} \; [\delta_{il} \delta_{kj} -
    {1\over N_c} \delta_{ij} \delta_{kl}]
\label{TaTa}
\end{equation}

\noindent for the generators in the fundamental representation
and the Fierz identity in $D=4$ for the product
$\gamma^{\mu} \otimes \gamma_{\mu} $ , one has a Lagrangian for the 
fermionic system

\begin{eqnarray}
{\cal L}_{fer}^{(n.a.)} \,& = & \,  
    {\bar \chi}_I i \gamma^\mu \partial_\mu  \chi_I  + {g \over M^{2}}
    \left[
    {\bar {\chi}}_I \chi_J {\bar {\chi}}_J \chi_I 
     - {\bar {\chi}}_I \gamma^5 \chi_J 
   {\bar {\chi}}_J \gamma^5 \chi_I \,\, \right.  \nonumber \\ 
    &  & \left.  +  {\bar {\chi}}_I \gamma^{\mu} \chi_J \,\,
    {\bar {\chi}}_J \gamma_{\mu} \chi_I  -
     {\bar {\chi}}_I \gamma^5 \gamma^{\mu} \chi_J \,\,
    {\bar {\chi}}_J \gamma_{\mu} \gamma^5 \chi_I \right] 
    \nonumber \\
    &  & + \, {\cal O} ({1\over N_c} , {M^2 \over \Lambda^2} ) \,
    + \, \lq\lq higher \,\,\, dimensional \,\,\, terms " \;,
\label{Lfna}
\end{eqnarray}

\noindent where the ${\cal O} (M^2/\Lambda^2)$ are due to terms 
proportional to $\epsilon$ coming from the change of fermionic variables 
in the action, the \lq\lq higher dimensional terms\rq \rq come from
corrections 
to the approximation used in the gauge field integration and there are 
\mbox{${\cal O} (1/N_c)$} terms from the identity (\ref{TaTa}) for the 
generators $T^a$ in the fundamental representation.

The fermionic model obtained at leading order in $1/N_c$ is an extension
of the Nambu-Jona-Lasinio (ENJL) model. This result had been proposed
previously~\cite{DW} as a conjecture on the behavior of a gauge 
theory at intermediate energies, compatible with the chiral symmetry 
of the gauge theory at leading order in $1/N_c$, which allows to 
calculate the low energy parameters~\cite{Bij}. 
In fact the fermionic model is a 
particular case with the constraint \mbox{$G_S = 4 G_V$} on the two
couplings of the general ENJL model~\cite{Bij}. This relation 
among couplings, considered previously as a perturbative estimate
(one gluon exchange) valid only at short distances, has been reobtained
in this work at the nonperturbative level as a consequence of the 
introduction of the effective fermionic field $\chi$. A redefinition
of the fermionic transformation (\ref{eff}) would lead to a
reformulation of the same fermionic system with different variables.
Then the constraint on the parameters of the ENJL model is a 
consequence of the dynamical content of the derivation of the
effective fermionic model and it goes beyond pure symmetry 
arguments at low energies. The same conclusion can be derived
from a comparison of the very particular form of the $1/N_c$
term ($-g^{2}/[2N_{c}M^{2}] {\bar {\chi}}_I \gamma^{\mu} \chi_I \,\,
{\bar {\chi}}_J \gamma_{\mu} \chi_J$) with the general form~\cite{Kle}
for the $1/N_c$ terms in an ENJL model. The particular form of
the \lq\lq higher dimensional terms\rq \rq in (\ref{Lfna}) would
only be required if one considers the effective fermionic theory
at energies close to the scale $M$ but in this range of energies
it is more convenient to consider the original gauge theory.

A strong phenomenological argument in favour of the validity of 
the derivation of the fermionic model presented in this work is 
that values for the couplings of the chiral Lagrangian in 
reasonable agreement with the experimental values, are obtained
from the ENJL model (including the case \mbox{$G_S = 4 G_V$)} in the case 
of QCD \mbox{($N_f = N_c = 3$)}. It would be interesting to go beyond the 
dominant term in the $1/N_c$ expansion in the derivation of the 
chiral Lagrangian from the fermionic model.

Besides the constraint among the two couplings of a general ENJL 
model~\cite{Bij}, or more precisely the very particular form for
the effective fermionic Lagrangian in (\ref{efL}) as compared with 
the most general form for the four-fermion interaction satisfying the 
symmetry requirements set by QCD~\cite{Kle}, another result of the 
derivation of the fermionic model from the gauge theory is the relation
 \mbox{$g/M^2 = 1/(\epsilon \Lambda^2)$} 
among the parameters ($g,M^2$) of the fermionic model, the ultraviolet 
cut-off of the gauge theory used in the calculation of the Jacobian of the 
fermionic transformation and the parameter $\epsilon$ appearing in the 
definition of the effective fermionic field. In order to determine 
the two parameters ($g,M^2$) a matching of the results of the effective
theory at $E < M$ with those of the gauge theory should be considered. 
In the case of QCD an alternative is to determine 
indirectly these parameters from a fit to the low energy parameters.

Although there are no essential differences at the level of the change 
of fermionic variables between the abelian and 
nonabelian cases it can happen that the generation of a dynamical mass
is a consequence of the growth at low energies of the gluonic 
self-interactions and then that an effective fermionic description is
only possible in the non-abelian case. Alternatively one can consider 
the possibility of spontaneous symmetry breaking and dynamical mass 
generation in the abelian case as a manifestation in a fermionic model 
of a strongly coupled QED phase.

To summarize, a new method to derive fermionic fields and a fermionic
effective Lagrangian for gauge theories has been presented. The main
idea is to identify a change of variables, sensitive to the 
ultraviolet domain of the theory, which shows the presence of an
intermediate scale $M$ in the theory. We assume that this scale can 
be interpreted as a dynamically generated mass scale and also that
below this scale a description of the system in terms of only 
fermionic fields is possible. When this idea is applied to QCD an 
ENJL model is obtained at leading order in $1/N_c$ whose low energy 
limit is compatible with the low energy properties as described 
through a chiral Lagrangian. It could be interesting to explore 
other possible applications of the change of variables introduced in 
this work considering different gauge groups, different representations 
for the matter fields and  dimension of space-time. Another possible 
extension of the present work can be based on different choices for 
the transformation (i.e. different choices for $\omega_{\mu\nu}$).
 
This work was partially supported by CICYT contract AEN 97-1680
(J.L.C., L.V.), Fondecyt-Chile 1950278, 1960229 and Dicyt-USACH (J.~G.). 
One of us (J.~G.) is a recipient of a John~S.~Guggenheim Fellowship.

\end{document}